\documentclass[aps,prb,twocolumn,superscriptaddress,showpacs,floatfix]{revtex4}
\usepackage{graphicx,epsfig,amsmath}

\newcommand{\mub}{\mbox{$\mu_{B}$}}
\newcommand{\kb}{\mbox{$k_{B}$}}

\newcommand{\etal}{\textit{et al.~}}
\newcommand{\etalComma}{\textit{et al.,}}

\newcommand{\AlxGaAs}[2]{\mbox{$\text{Al}_{#1}\text{Ga}_{#2}\text{As}$}}
\newcommand{\Vds}{\mbox{$\text{V}_{\text{ds}}$}}
\newcommand{\Vdsstar}{\mbox{$\text{V}_{\text{ds}}^*$}}
\newcommand{\Bstar}{\mbox{$\text{B}_{||}^*$}}
\newcommand{\Tstar}{\mbox{$\text{T}^*$}}
\newcommand{\Vg}{\mbox{$\text{V}_{\text{g}}$}}
\newcommand{\Vr}{\mbox{$\text{V}_{\text{r}}$}}
\newcommand{\Vt}{\mbox{$\text{V}_{\text{t}}$}}
\newcommand{\Vb}{\mbox{$\text{V}_{\text{b}}$}}
\newcommand{\didv}{\mbox{dI/d\Vds}}

\newcommand{\Bpar}{\mbox{$\text{B}_{||}$}}

\newcommand{\subfig}[2]{Fig.~\ref{fig:#1}(#2)}
\newcommand\degrees[1]{\ensuremath{#1^\circ}}

\begin{document}

\title{Two-stage Kondo effect in a four-electron artificial atom}

\author{G. Granger}	
	\affiliation{Department of Physics, Massachusetts Institute of Technology, Cambridge, Massachusetts 
02139}
\author{M. A. Kastner} 
	\email{mkastner@mit.edu}
	\affiliation{Department of Physics, Massachusetts Institute of Technology, Cambridge, Massachusetts 
02139}

\author{Iuliana Radu} 
	\affiliation{Department of Physics, Massachusetts Institute of Technology, Cambridge, Massachusetts 
02139}

\author{M. P. Hanson}
	\affiliation{Materials Department, University of California, Santa Barbara 93106-5050}
\author{A. C. Gossard}
	\affiliation{Materials Department, University of California, Santa Barbara 93106-5050}


\begin{abstract}

An artificial atom with four electrons is driven through a singlet-triplet transition by varying the confining potential. In the triplet, a Kondo peak with a narrow dip at drain-source voltage \Vds=0 is observed. The low energy scale $\Vdsstar$ characterizing the dip is consistent with predictions for the two-stage Kondo effect. The phenomenon is studied as a function of temperature T and magnetic field $\Bpar$, parallel to the two-dimensional electron gas. The low energy scales $\Tstar$ and $\Bstar$ are extracted from the behavior of the zero-bias conductance and are compared to the low energy scale $\Vdsstar$ obtained from the differential conductance. Good agreement is found between $\kb\Tstar$ and $|g|\mub\Bstar$, but $e\Vdsstar$ is larger, perhaps because of nonequilibrium effects.

\end{abstract}

\pacs{73.23.-b, 73.63.Kv, 73.23.Hk, 75.20.Hr}

\maketitle


\section{Introduction}

In recent years, much attention has been paid to experiments on the Kondo effect in nanostructures, particularly to the spin-1/2 Kondo effect in single-electron transistors.\cite{goldhaber1998:NatureKondo,cronenwett1998:ScienceKondo} The Kondo effect arises when virtual transitions from a spin degenerate level on a quantum dot to the continuum states in the reservoirs cause the formation of the many-body Kondo singlet. This entangled state leads to an enhanced conductance at low temperature. 

The presence of ground states with spin $S>1/2$ has been detected in semiconductor quantum dots \cite{schmid2000:triplet} and in carbon nanotubes. \cite{liang2002:nanotube_triplet} Transitions between different spin states have also been studied. Specifically, transitions between a spin singlet and a spin triplet ground state have been observed in vertical and lateral quantum dots and carbon nanotubes. \cite{su1994:DBRTS_ST,sasaki2000:verticalST,vdwiel2002:lateralST,zumbuhl2004:cotunnelingspec,nygard2000:nanotube_ST} In these experiments a magnetic field has been used to change the energy spacing between two spatial orbitals, and the triplet becomes the ground state when the energy spacing becomes smaller than the Hund's-rule exchange. Kyriakidis \etal\cite{kyriakidis2002:voltage-tunableST} noted that, at fixed magnetic field, deforming the dot with a side gate voltage could also drive the singlet-triplet transition, and Kogan \etal \cite{kogan2003:ST_zeroB} have observed such transitions in the Kondo regime at zero magnetic field in this way. In the absence of a magnetic field, the ground state of the quantum dot is degenerate if the total spin on the dot is different from zero, and the enhanced conductance from the Kondo effect is observed. One can then identify the regions of nonzero spin from the sharp Kondo features in the differential conductance.\cite{kogan2003:ST_zeroB} A singlet-triplet transition tuned by asymmetric gate voltages has also been reported for a quantum ring in the Coulomb-blockade regime at zero magnetic field.\cite{fuhrer2003:ringST}

There have been numerous theoretical studies of singlet-triplet transitions in vertical dots
\cite{pustilnik2001:theory_verticalST,izumida2001:theory_verticalST,eto2002:theory_verticalST}, lateral dots \cite{pustilnik2001:no_excited_state_theoryST,hofstetter2002:first_theoryST,pustilnik2003:theoryST_source_eq8}, and coupled dots.\cite{golovach2003:coupled_dot_theoryST} In particular, Hofstetter and Zarand \cite{hofstetter2004:theoryST} have studied the singlet-triplet crossover in a lateral quantum dot with an even number of electrons. They predict a two-stage Kondo effect in the triplet region resulting from the presence of two Kondo energy scales, which can be very different. They have also provided predictions for the zero-bias conductance of the dot as a function of a magnetic field parallel to the two-dimensional electron gas (2DEG) for both the singlet and the triplet regions as well as for the crossover region. In particular, this dependence is predicted to be non-monotonic on the triplet side. They expect these equilibrium predictions to be qualitatively applicable to the description of both the temperature dependence of the zero-bias conductance and the drain-source voltage dependence of the differential conductance, even though the latter explores nonequilibrium physics.

When analyzing lateral quantum dots near a singlet-triplet degeneracy, it is important to determine how many modes (sometimes called ``channels'') take part in the Kondo screening of the two relevant orbitals. In the Coulomb-blockade regime, the quantum point contacts, which form the tunnel barriers, are pinched off, so each barrier can be viewed as a waveguide with only one propagating mode. This is different from vertical dots, where the geometry allows several propagation modes from each lead. If each of the two orbitals of the lateral dot couples to its own linear combination of the two modes, a triplet can be screened entirely and form a many-body singlet with the leads. For instance, assuming the two orbitals have different parity, the symmetric combination of the two modes from the leads will tend to screen the symmetric orbital, whereas the antisymmetric combination will tend to screen the antisymmetric one. There is a Kondo temperature associated with each of the two screening processes. In general, these two Kondo temperatures are expected to be different, so the many-body singlet forms in two steps, hence the name ``two-stage Kondo effect.'' On the other hand, it is possible that one of the two combinations of modes decouples, in which case the triplet can only be partially screened, resulting in a many-body doublet. The theory of Ref.~\cite{hofstetter2004:theoryST} makes predictions for various degrees of coupling between the orbitals and the two combinations of modes.

In this article we present a set of experiments designed to test some of these predictions. We create an artificial atom containing four electrons and induce the transition from singlet to triplet by changing the confining potential.

Our paper is divided as follows. In section~\ref{sec:exp_details}, details about the device studied and the measurement setup are provided. In section~\ref{sec:results}, we present differential conductance data. The quantum dot is first emptied of all its electrons, and, once four electrons are added, the system is driven through a singlet-triplet crossover by deforming the dot potential with a side gate. The dependences of the conductance on drain-source voltage, temperature, and parallel magnetic field reveal that two different energy scales characterize the triplet, as expected for the two-stage Kondo effect. A simple quadratic fit is used in section~\ref{sec:analysis} to extract the low energy scales in the zero-bias conductance data from the parallel magnetic field, temperature, and drain-source voltage dependences. It is the first time that the low energy scale is measured using a parallel magnetic field. The results are compared and discussed in section~\ref{sec:discussion}, where it is found that the low energy scale measured from the drain-source voltage dependence is higher than those measured from the temperature and parallel magnetic field dependences. We ascribe this difference to nonequilibrium effects.

\section{Experimental details}
\label{sec:exp_details}

\begin{figure}[bt]
\setlength{\unitlength}{1cm}
\begin{center}
\begin{picture}(8,4.5)(0,0)
\put(1.5,0){\includegraphics[width=5cm, keepaspectratio=true]{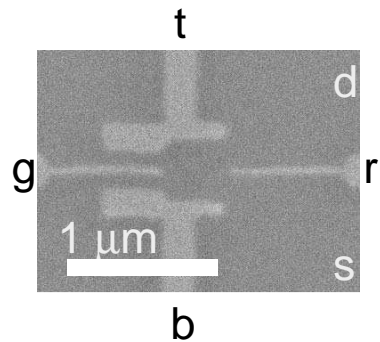}}
\end{picture}
\end{center}
\caption{Electron micrograph of a device nominally identical to that under study. The four gate electrodes are labeled with the indices g, r, t, and b. The drain and source are indicated by d and s, respectively. Lead d is biased, while lead s stays at virtual ground. The bar corresponds to 1 $\mu$m.}
\label{fig:Dot_Micrograph}
\end{figure}

The device studied is an artificial atom defined by electrostatic means in an \AlxGaAs{0.3}{0.7}/GaAs heterostructure grown by Molecular Beam Epitaxy, in which the 2DEG is located 110~nm below the surface. Magnetotransport at dilution refrigerator temperatures reveals that the density of the 2DEG is $2.2\times 10^{15}$~m$^{-2}$ and the mobility is 64~m$^2$/Vs.  We anneal NiGeAu pads to make ohmic contacts to the 2DEG. Large TiAu gate electrodes are patterned by photolithography, and submicron CrAu electrodes are defined by electron-beam lithography (see Fig.~\ref{fig:Dot_Micrograph} for their geometry).

Applying negative voltages to the gate electrodes creates an artificial atom of less than 400 nm diameter thanks to the depletion that occurs near the gates. Two remaining portions of 2DEG that are well coupled to the artificial atom and that lead to ohmic contacts are used as the drain and the source (labeled d and s on Fig.~\ref{fig:Dot_Micrograph}) for transport measurements. The number of electrons on the artificial atom is controlled by the plunger gate voltage $\Vg$ applied to electrode g, while the voltages (\Vr,\Vt,\Vb) on electrodes r, t, and b (see Fig.~\ref{fig:Dot_Micrograph}) are used for the confinement of the electrons and are usually held fixed. By adding a peak-to-peak modulation of 2.5 $\mu$V to the DC voltage $\Vds$ applied between the drain and the source, the differential conductance $\didv$ is measured with a current amplifier and a lock-in amplifier.

The device is placed inside a dilution refrigerator with a base mixing chamber temperature below 10~mK and an electron base temperature of 30~mK. The sample is aligned so the magnetic field is parallel to the 2DEG to within \degrees{0.5}. Orbital effects can be neglected up to magnetic fields such that the magnetic length due to the unwanted perpendicular component becomes comparable to the dot size. For our alignment precision and dot size, orbital effects are expected to become significant only above \Bpar=3 T.

\section{Results}
\label{sec:results}

Devices with electrode patterns similar to those in Fig.~\ref{fig:Dot_Micrograph} have been reported to be useful in the process of emptying a lateral dot of all its electrons.\cite{ciorga2000:n=0} Placing the drain and the source on the same side of the dot allows one to keep the couplings large enough to do transport measurements in the Kondo regime, even though most electrodes are biased very negatively.

\begin{figure}[hbt]
\setlength{\unitlength}{1cm}
\begin{center}
\begin{picture}(8,9.5)(0,0)
\put(0,9.5){(a)}
\put(0,5){\includegraphics[width=8cm, keepaspectratio=true]{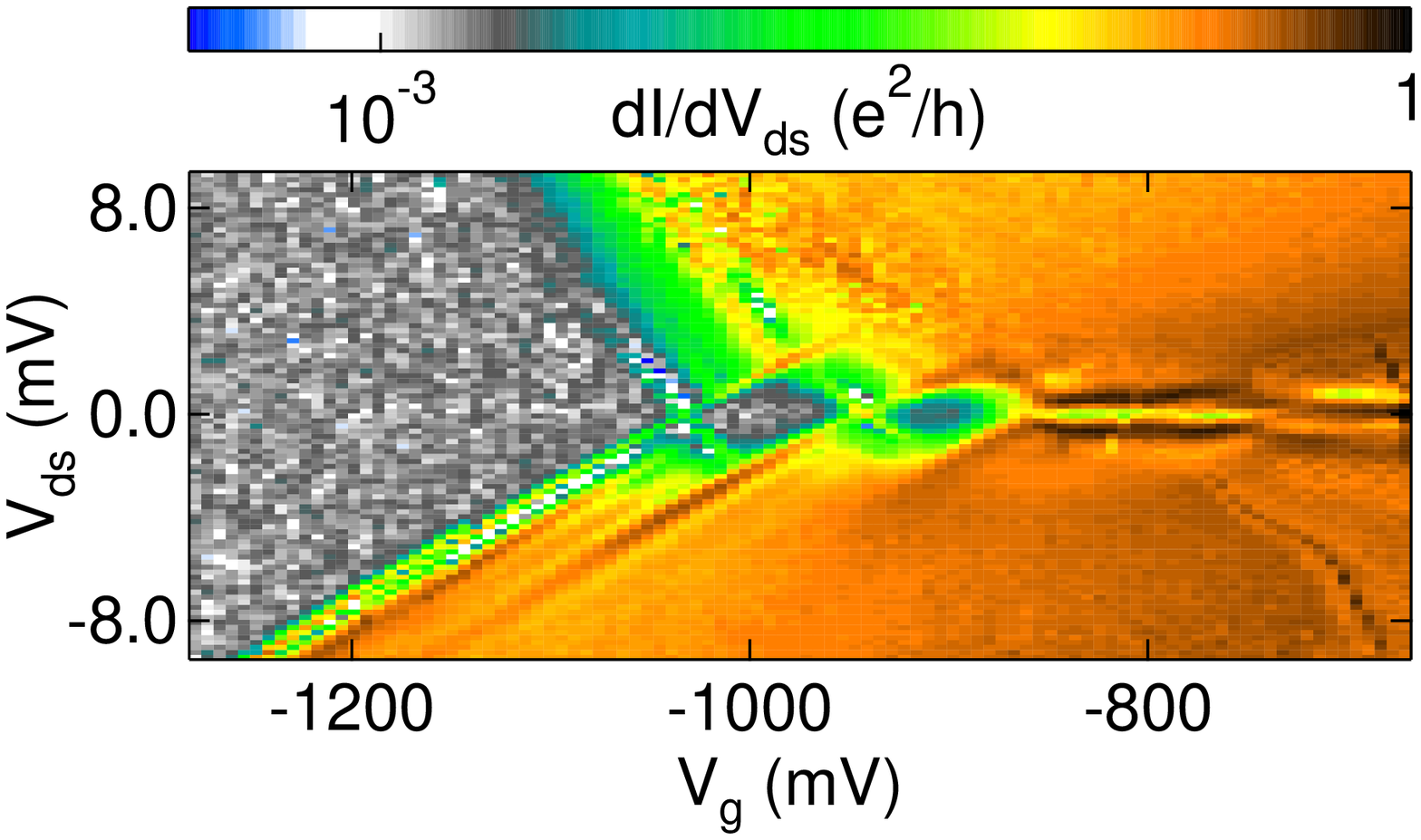}}
\put(0,4.5){(b)}
\put(0,0){\includegraphics[width=8cm, keepaspectratio=true]{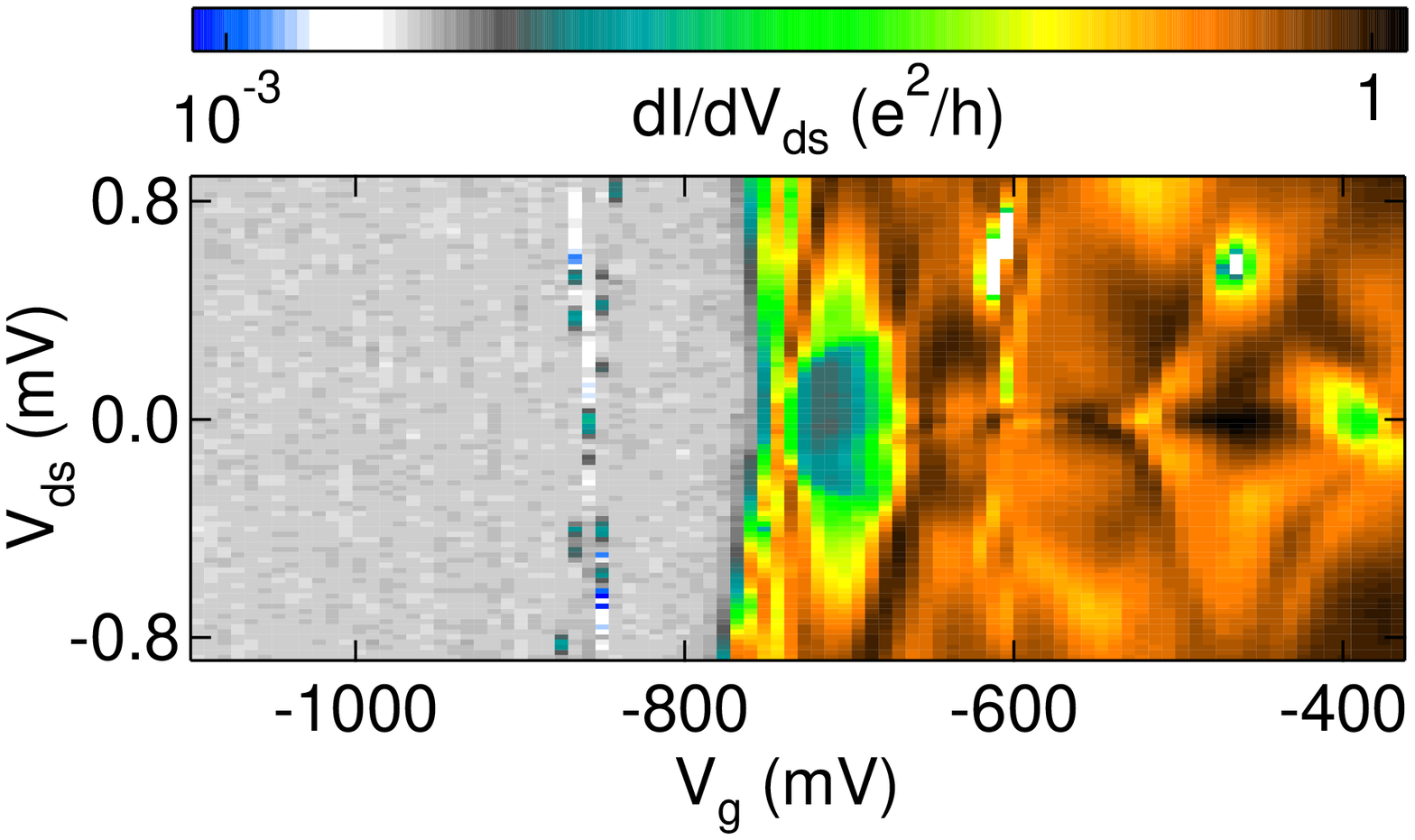}}
\end{picture}
\end{center}
\caption{(a) (color online) Logarithmic map of $\didv$ in the $\Vds$-$\Vg$ plane showing that the artificial atom can be depleted of all its electrons. The constricting electrode voltages are at (\Vr,\Vt,\Vb)=(-762,-904,-862)~mV.  (b) (color online) Logarithmic $\didv$ map in the $\Vds$-$\Vg$ plane from a separate cool down for (\Vr,\Vt,\Vb)=(-762,-904,-862)~mV in a narrower range of \Vds. The first charging feature is at \Vg=-858 mV. All data from this point on were taken in the same cool down.}
\label{fig:Emptying}
\end{figure}

Figure~\ref{fig:Emptying}(a) contains a differential conductance map that proves that the artificial atom can be emptied of all its electrons. The last Coulomb charging peak is clearly seen at \Vg=-1033~mV. The charging features at finite $\Vds$ on the left of this peak never close back again to form a diamond, even if $\Vds$ is increased beyond 8 mV of either polarity. Such large values of $\Vds$ are much beyond the charging energy of the first Coulomb-blockade valley, which is $U_1=(2.0\pm0.2)$~meV.

The remaining data discussed in this paper are from a different cool down. Once the dot has been emptied, it is easy to keep track of the electron number when tuning the voltages on various electrodes to search for a region of interest. The $\didv$ map in Fig.~\ref{fig:Emptying}(b) shows a range of $\Vg$ where the first few electrons are added. The amplitude of the first Coulomb charging peak, located at \Vg=-858~mV, is now very small compared to that of the subsequent ones. The dot rapidly becomes fairly open, as seen by the widening of the charging features as $\Vg$ is increased. We have further checked that the first electron is added at \Vg=-858~mV by opening the constrictions with $\Vr$ and closing and symmetrizing with both $\Vt$ and $\Vb$ to make the first peak grow as tall as possible. This would make any features at more negative $\Vg$ more easily observable.

\begin{figure}[hbt!]
\setlength{\unitlength}{1cm}
\begin{center}
\begin{picture}(8,11)(0,0)
\put(0,3.5){\includegraphics[width=8cm, keepaspectratio=true]{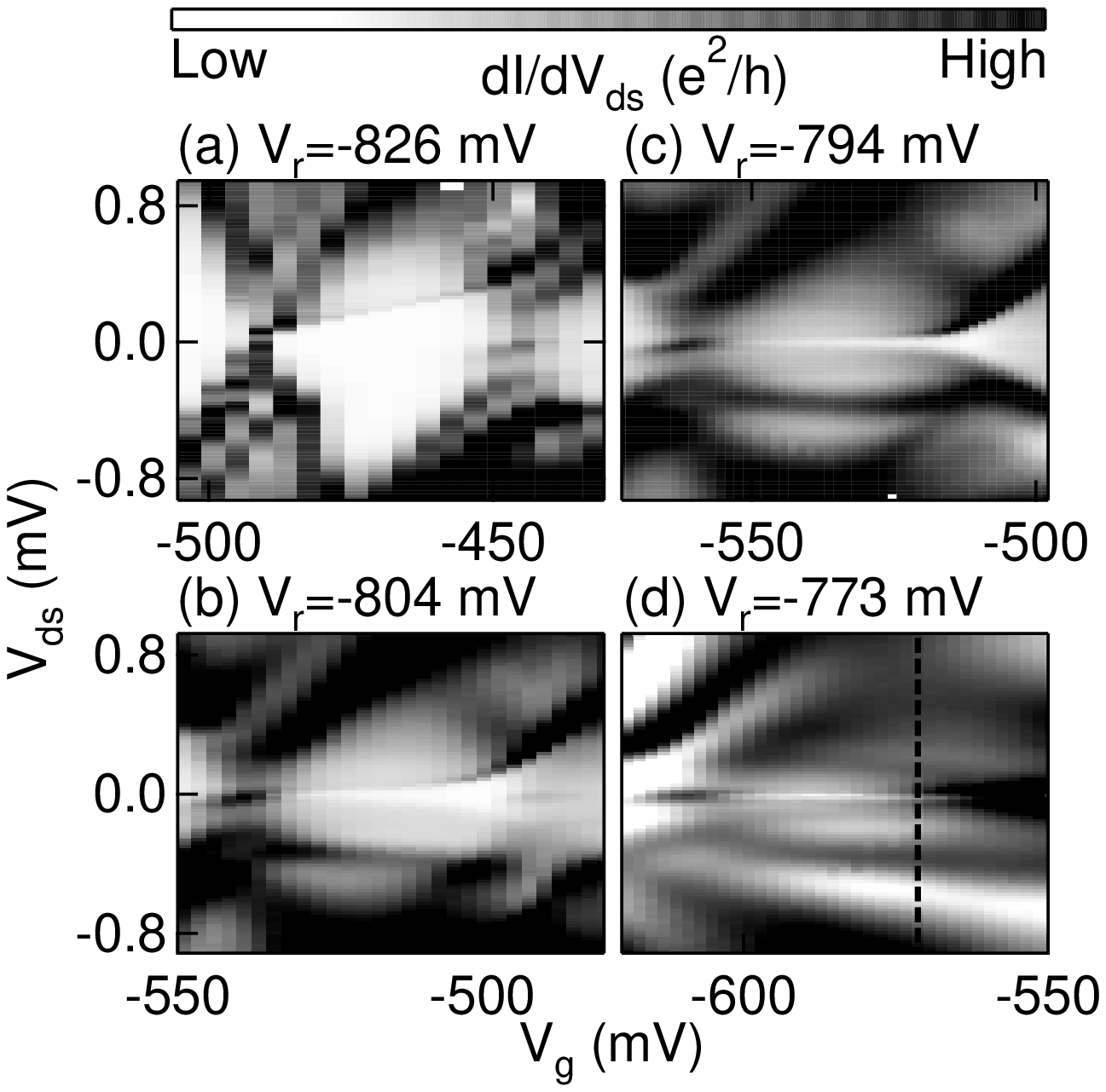}}
\put(4.2,0){\includegraphics[width=3.87cm, keepaspectratio=true]{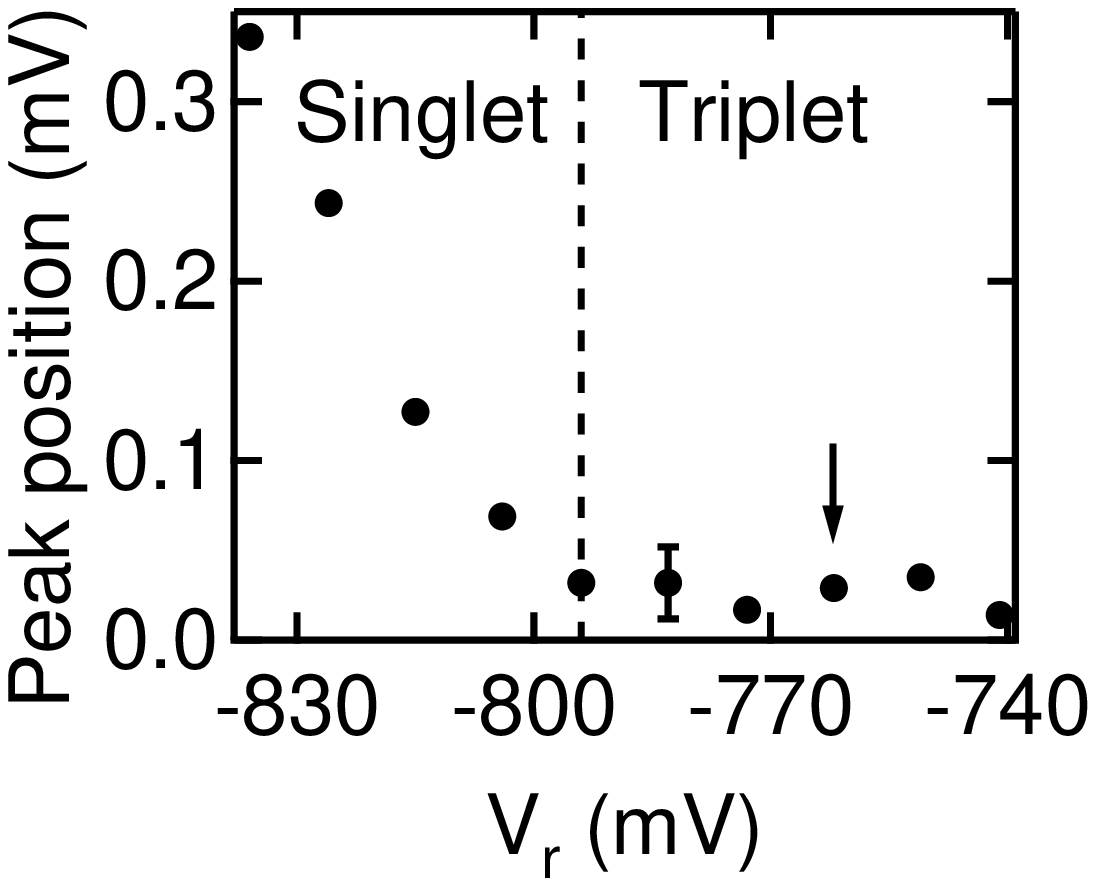}}
\put(0.1,0.1){\includegraphics[width=3.85cm, keepaspectratio=true]{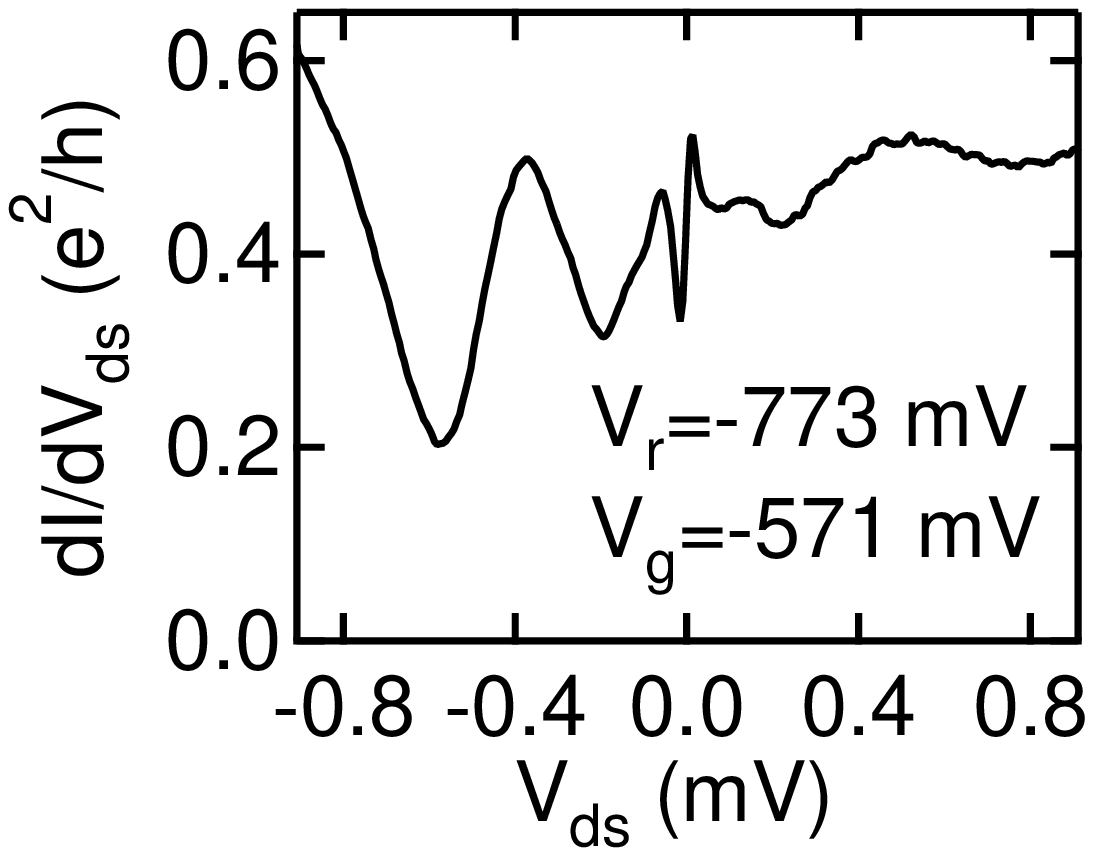}}
\put(1.1,3.2){(e)}
\put(5,3.2){(f)}
\end{picture}
\end{center}
\caption{Changing $\Vr$ to induce a ground state singlet-triplet transition. The displayed range of $\Vg$ is adjusted from (a) to (d) to take into account the capacitive shift induced by changes in \Vr. The gray scale is adjusted to enhance the relevant features, but white is low $\didv$ and black is high. (\Vt,\Vb)=(-904,-862)~mV in all the subfigures. (a) \Vr=-826~mV. In the middle diamond, the ground state is the singlet. The gate-voltage dependent, Kondo-enhanced inelastic cotunneling threshold for the excited triplet state is visible only at positive \Vds. White (black) corresponds to 0 (0.3) $e^2/h$. (b)  \Vr=-804~mV. The threshold to make the triplet excited state is now much closer to \Vds=0 and can be seen at either polarity. White (black) corresponds to 0 (0.45) $e^2/h$. (c) \Vr=-794~mV. The threshold for the triplet excitation has closed in even more toward \Vds=0. White (black) corresponds to 0 (0.55) $e^2/h$. (d) \Vr=-773~mV. The threshold has evolved into a broad Kondo peak with a dip in the middle. The new ground state is the triplet in the two-stage Kondo regime. White (black) corresponds to 0.2 (0.6) $e^2/h$. (e) Differential conductance in the two-stage Kondo regime taken along the dashed line at $\Vg=-571$~mV in (d). (f) $\Vr$ dependence of the position of the peak near the inelastic cotunneling threshold at positive $\Vds$ in the middle of the N=4 diamond (28~mV to the right of the left conductance peak). The dashed line indicates the value of $\Vr$ where the ground state singlet-triplet transition occurs. A typical error bar is shown. The arrow indicates $\Vr=-762$~mV.}
\label{fig:Singlet-triplet}
\end{figure}

In order to study the two-stage Kondo effect, one needs to drive the dot into a triplet ground state. It is known from Ref.~\cite{kogan2003:ST_zeroB} that the deformation caused by varying the confining potential can induce a singlet-triplet transition when two orbitals come close enough to each other. Figure~\ref{fig:Singlet-triplet} shows the evolution of a region of $\didv$ as $\Vr$ is made less negative. The central Coulomb-blockade diamond corresponds to the case with N=4 electrons on the dot. In Fig.~\ref{fig:Singlet-triplet}(a), \Vr=-826~mV. In spite of the relatively low resolution of this first subfigure, the inelastic cotunneling threshold, which varies with gate voltage, is visible at positive $\Vds$ in the central diamond. This threshold is believed to result from internal excitations of the 4-electron artificial atom from a singlet ground state to a triplet excited state. When $\Vr$ is increased to -804~mV, as seen in Fig.~\ref{fig:Singlet-triplet}(b), the excited state feature is closer to \Vds=0 and can be seen for either sign of \Vds. As expected, the change in $\Vr$ also causes a capacitive shift, which explains why the range of $\Vg$ is moved to more negative values from one subfigure to the next.  As $\Vr$ is increased further to -794~mV and -773~mV (Figs.~\ref{fig:Singlet-triplet}(c) and (d)), the excited state evolves into a narrow dip at zero bias inside of a broader peak (see Fig.~\ref{fig:Singlet-triplet}(e) for an example of a trace at fixed $\Vg$), typical of a triplet ground state. The additional broad features near $\pm0.4$~mV probably correspond to higher-energy states with spin.

Figure~\ref{fig:Singlet-triplet}(f) illustrates that changing $\Vr$ allows the triplet to become the ground state. The position of the peaks near the inelastic cotunneling features at $\Vds>0$ is extracted near the middle of diamonds such as those in Fig.~\ref{fig:Singlet-triplet}(a) to (d). The peak position is plotted as a function of $\Vr$ over the range from -836~mV to -741~mV, and it is found to decrease before saturating. This is very similar to the behavior reported by Kogan \etal\cite{kogan2003:ST_zeroB} The ground state transition from a singlet to a triplet occurs near $\Vr=-794$~mV, i.e.~where the behavior of the peak position becomes a constant slightly greater than zero. The Coulomb charging peak spacings observed in zero-bias conductance measurements as a function of $\Vg$ remain constant as $\Vr$ is increased in small steps over the range shown in Fig.~\ref{fig:Singlet-triplet}(f), indicating that we are indeed studying a single dot. In the rest of this paper, the situation where \Vr=-762~mV, indicated by an arrow in Fig.~\ref{fig:Singlet-triplet}(f), will be studied in detail.

\begin{figure}[hbt]
\setlength{\unitlength}{1cm}
\begin{center}
\begin{picture}(8,10.5)(0,0)
\put(0.1,10){(a)}
\put(0,0){\includegraphics[width=8.0cm, keepaspectratio=true]{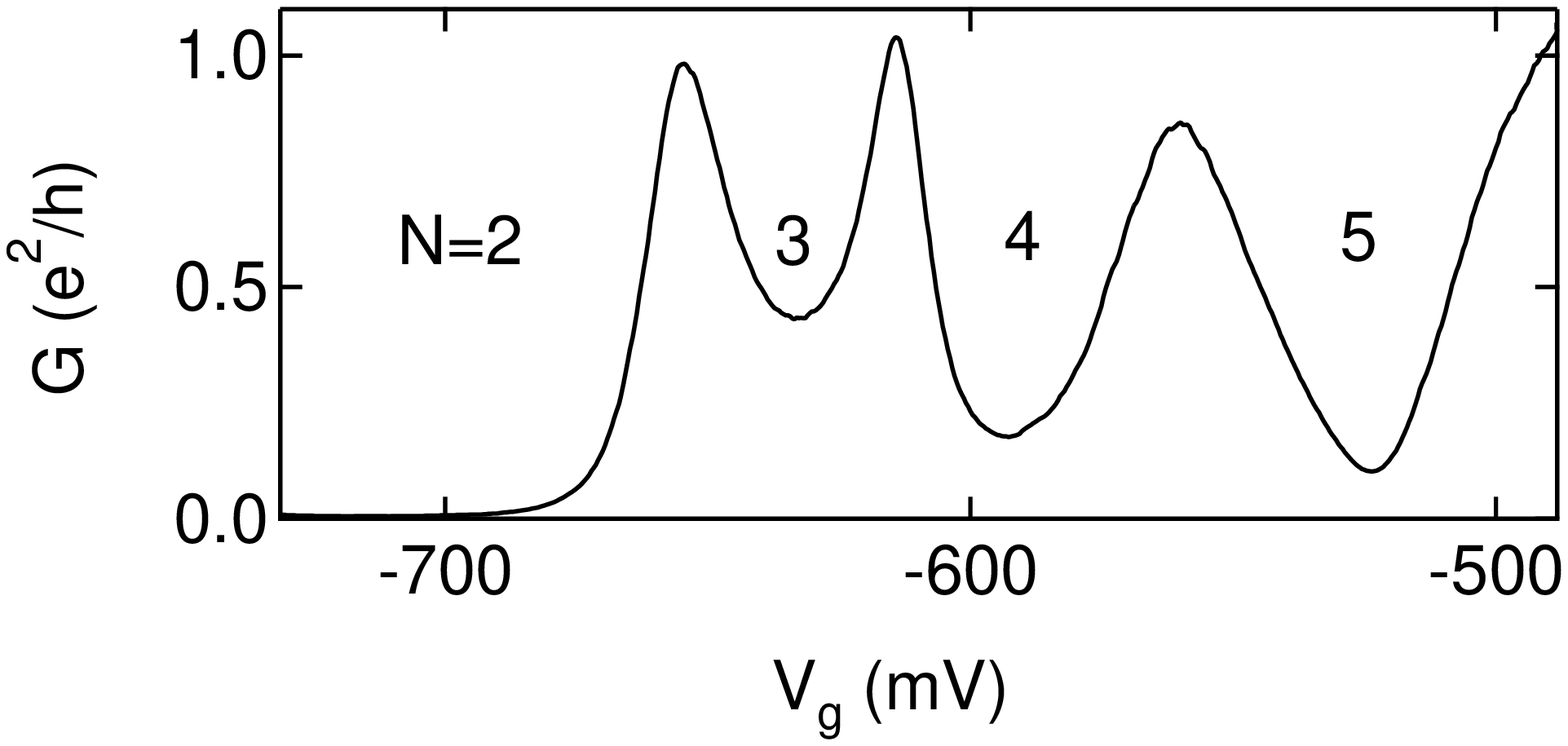}}
\put(0.1,3.5){(b)}
\put(0.15,3.9){\includegraphics[width=7.9cm, keepaspectratio=true]{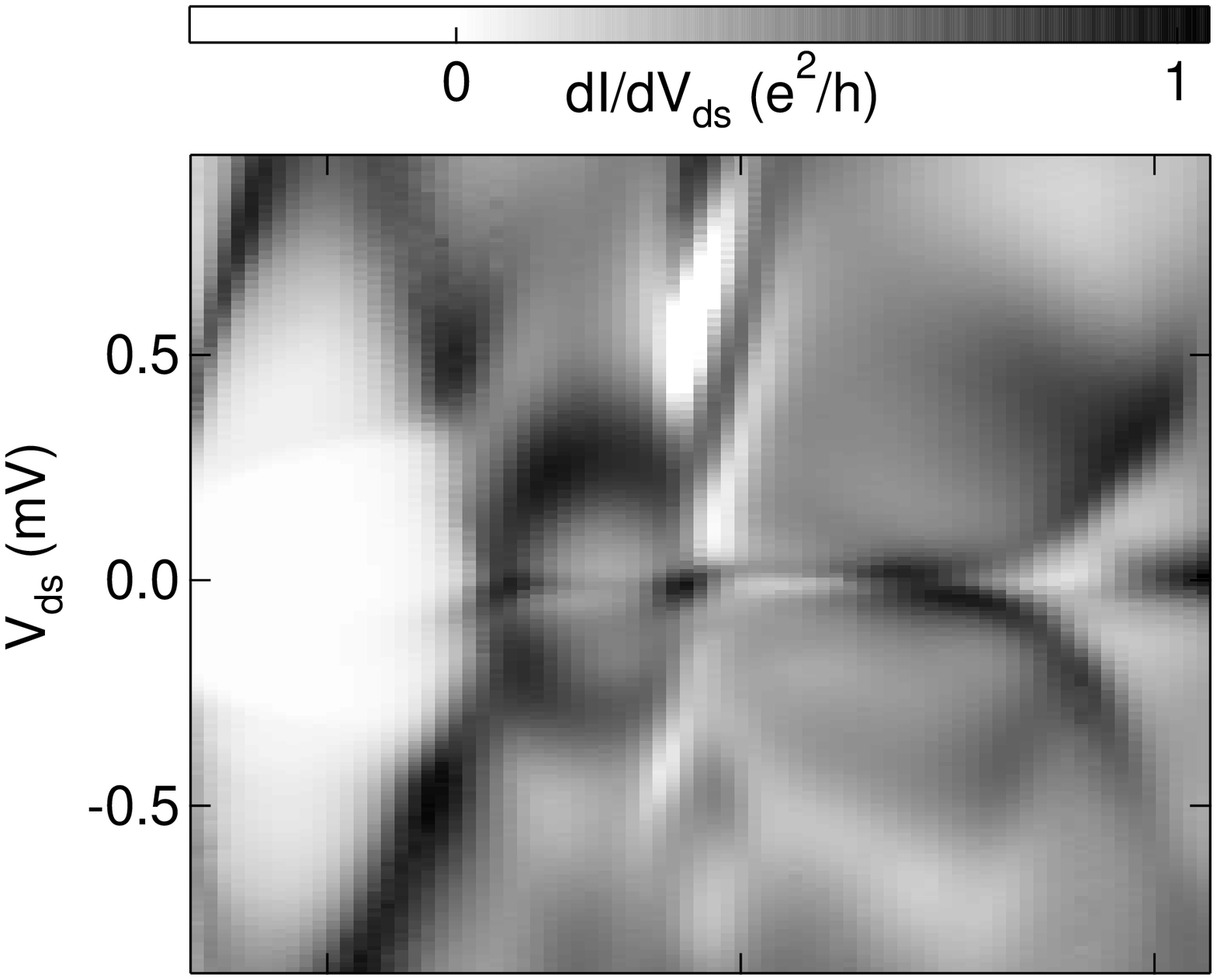}}
\end{picture}
\end{center}
\caption{(a) $\didv$ map in the $\Vds$-$\Vg$ plane in the few-electron regime starting with the diamond for N=2, near \Vg=-700~mV. The constricting electrode voltages are at (\Vr,\Vt,\Vb)=(-762,-904,-862) mV. A \Vg-dependent inelastic cotunneling threshold is observed in the N=2 diamond. The N=3 diamond shows the ordinary spin-1/2 Kondo effect. The N=4 and N=5 diamonds show a more complex behavior. (b) Zero-bias conductance G as a function of \Vg. These data can be compared directly with a trace of the $\didv$ map in (a) taken at \Vds=0. The number of electrons on the dot in each Coulomb-blockade valley is indicated.}
\label{fig:LargeDiamonds}
\end{figure}

Figure~\ref{fig:LargeDiamonds}(a) shows the differential conductance $\didv$ as a function of $\Vds$ and $\Vg$ for \Vr=-762~mV at zero magnetic field and base temperature. The left diamond corresponds to the Coulomb-blockade valley with an electron number N=2. One can see inelastic cotunneling thresholds at $\Vds\sim\pm0.25$~mV that vary with gate voltage. The size of the cotunneling gap can be used to determine the energy separation between the singlet ground state and the triplet excited state.\cite{kogan2003:ST_zeroB} In the next diamond, where N=3, there is a very sharp feature at \Vds=0 that we assign to the spin-1/2 Kondo effect. There is also an unusually strong inelastic cotunneling feature away from zero bias, which must result from higher-lying states. The zero-bias conductance data taken under the same conditions as Fig.~\ref{fig:LargeDiamonds}(a) is presented in \subfig{LargeDiamonds}{b}. The valleys with N=4 and N=5 are difficult to distinguish on the $\didv$ plot, but it is very easy to locate all the charging peaks in this neighborhood by looking at the behavior of G(\Vg). It is unclear why the charging features at nonzero $\Vds$ are very faint in \subfig{LargeDiamonds}{a} for the Coulomb-charging peak between the valleys with N=4 and N=5. We focus our attention to the Coulomb-blockade valley with N=4.

\begin{figure}[hbt] 
\setlength{\unitlength}{1cm}
\begin{center}
\begin{picture}(8,6)(0,0)
\put(0,5.5){(a)}
\put(0,0.1){\includegraphics[width=3.3cm, keepaspectratio=true]{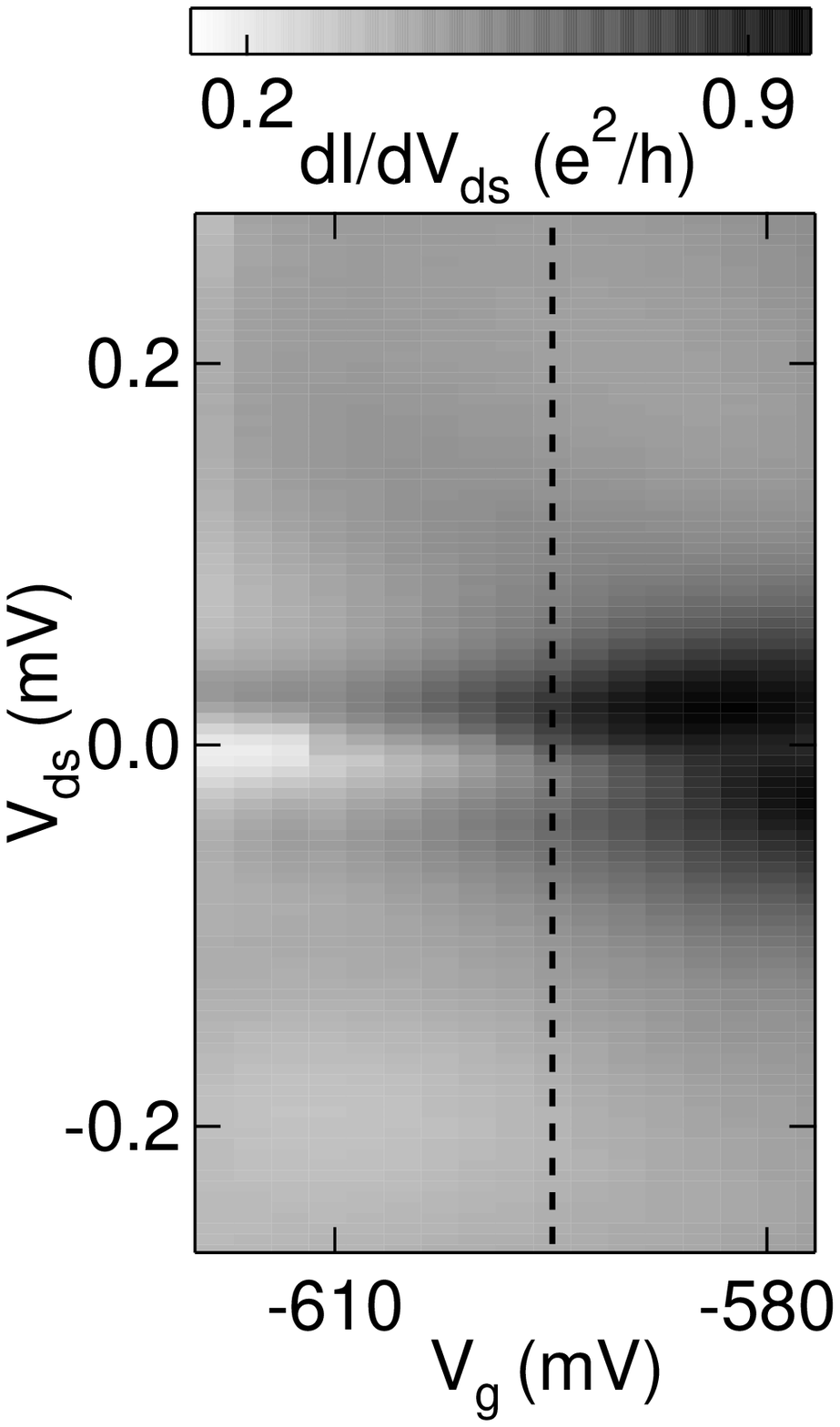}}
\put(3.5,5.5){(b)}
\put(3.5,0){\includegraphics[width=3.9cm, keepaspectratio=true]{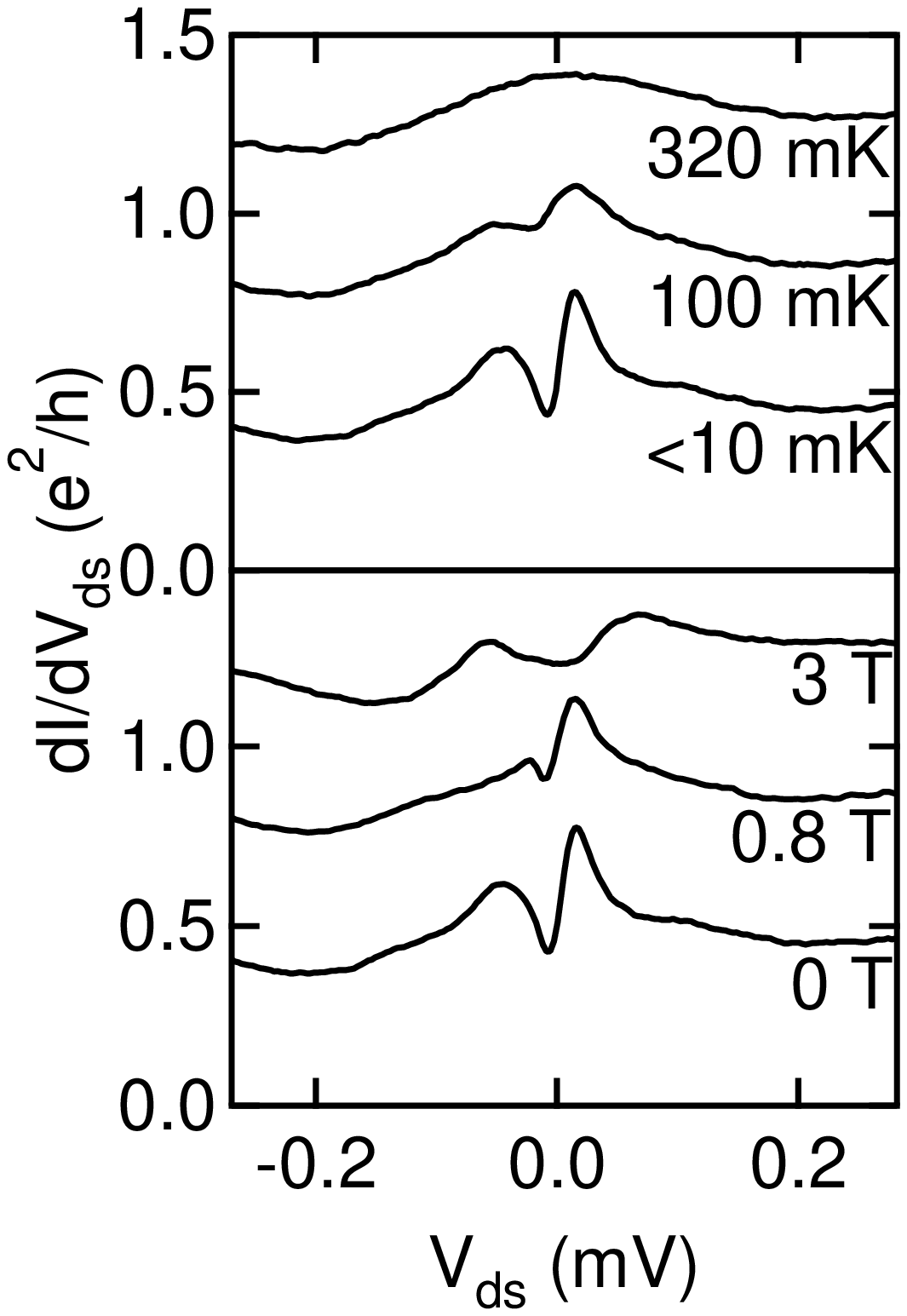}}
\end{picture}
\end{center}
\caption{(a) $\didv$ map in the $\Vds$-$\Vg$ plane for N=4 when (\Vr,\Vt,\Vb)=(-762,-904,-862)~mV at $\Bpar=0$~T and base temperature. The suppression in the zero-bias peak in $\didv$ is ascribed to a two-stage Kondo effect in the triplet ground state. Because of hysteresis after sweeping to very negative \Vg, the features have moved to the left by 19~mV compared to Fig.~\ref{fig:LargeDiamonds}(b). (b) The top panel shows $\didv$ as a function of $\Vds$ at \Vg=-595 mV (along the dashed line in (a)) at three different mixing chamber temperatures. T from bottom up: base temperature ($<$ 10), 100, and 320 mK. The bottom panel shows the $\Bpar$ evolution at \Vg=-595 mV. $\Bpar$ from bottom up: 0, 0.8, and 3.0 T. There is a 0.4 $e^2/h$ offset between the curves for clarity.}
\label{fig:Diamond4}
\end{figure}

In \subfig{Diamond4}{a}, we show an expanded version of the $\didv$ map for the Coulomb-blockade valley with N=4. For each trace of $\didv$ versus $\Vds$ at fixed $\Vg$ in this valley, a peak is found with a small dip at zero bias. An example of such a trace is shown for \Bpar=0~T and base temperature in Fig.~\ref{fig:Diamond4}(b). In order to investigate whether this situation indeed describes the two-stage Kondo effect, both the temperature and the parallel magnetic field dependences for the conductance throughout this valley have been measured. 

Measurements of $\didv$ versus $\Vds$ at the constant plunger gate voltage of \Vg=-595~mV are shown in the top panel of Fig.~\ref{fig:Diamond4}(b), for three temperatures. At base temperature, a clear zero-bias dip is observed in the wider zero-bias peak. A fit to a sum of two Lorentzians with amplitudes of opposite signs added to a constant background provides a reasonable fit to the data (not shown), allowing a rough estimate for the two Kondo scales involved. The wider Lorentzian has a width of order 100~$\mu$eV, while the narrower one has a width of order 40~$\mu$eV. The zero-bias dip has nearly disappeared for the \mbox{T}=100~mK trace, and the broadening of the underlying peak becomes obvious by \mbox{T}=320~mK. 

The bottom panel of \subfig{Diamond4}{b} shows $\didv$ curves for three different values of \Bpar, taken at base temperature. At \Bpar=0.8~T, the dip at zero bias is less pronounced than at \Bpar=0~T. The Zeeman splitting of the central peak is obvious at \Bpar=3~T.

\begin{figure}[hbt]
\setlength{\unitlength}{1cm}
\begin{center}
\begin{picture}(8,11.2)(0,0)
\put(0,10.7){(a)}
\put(0,5.53){\includegraphics[width=4.05cm, keepaspectratio=true]{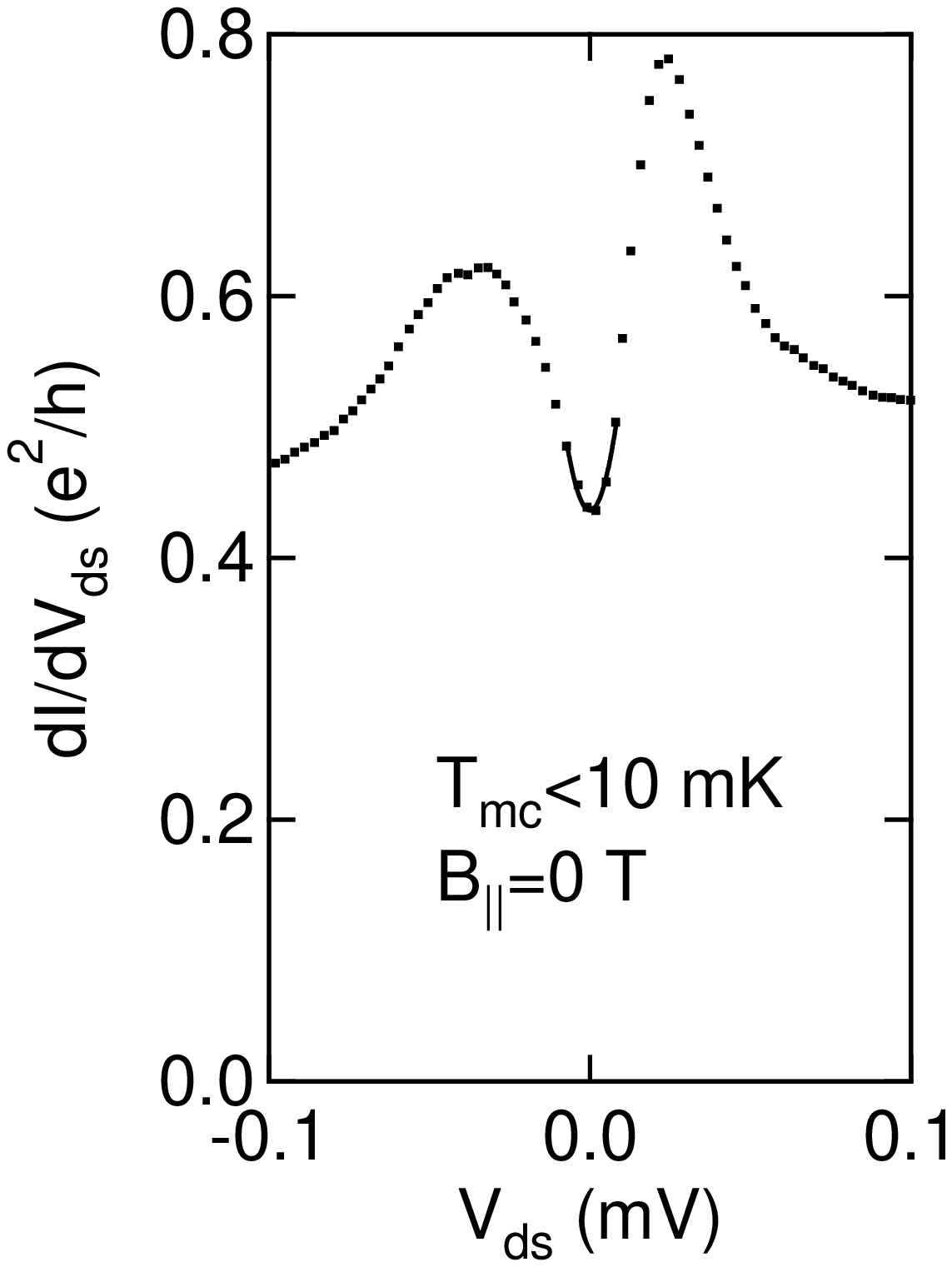}}
\put(4,10.7){(b)}
\put(4,5.5){\includegraphics[width=3.77cm, keepaspectratio=true]{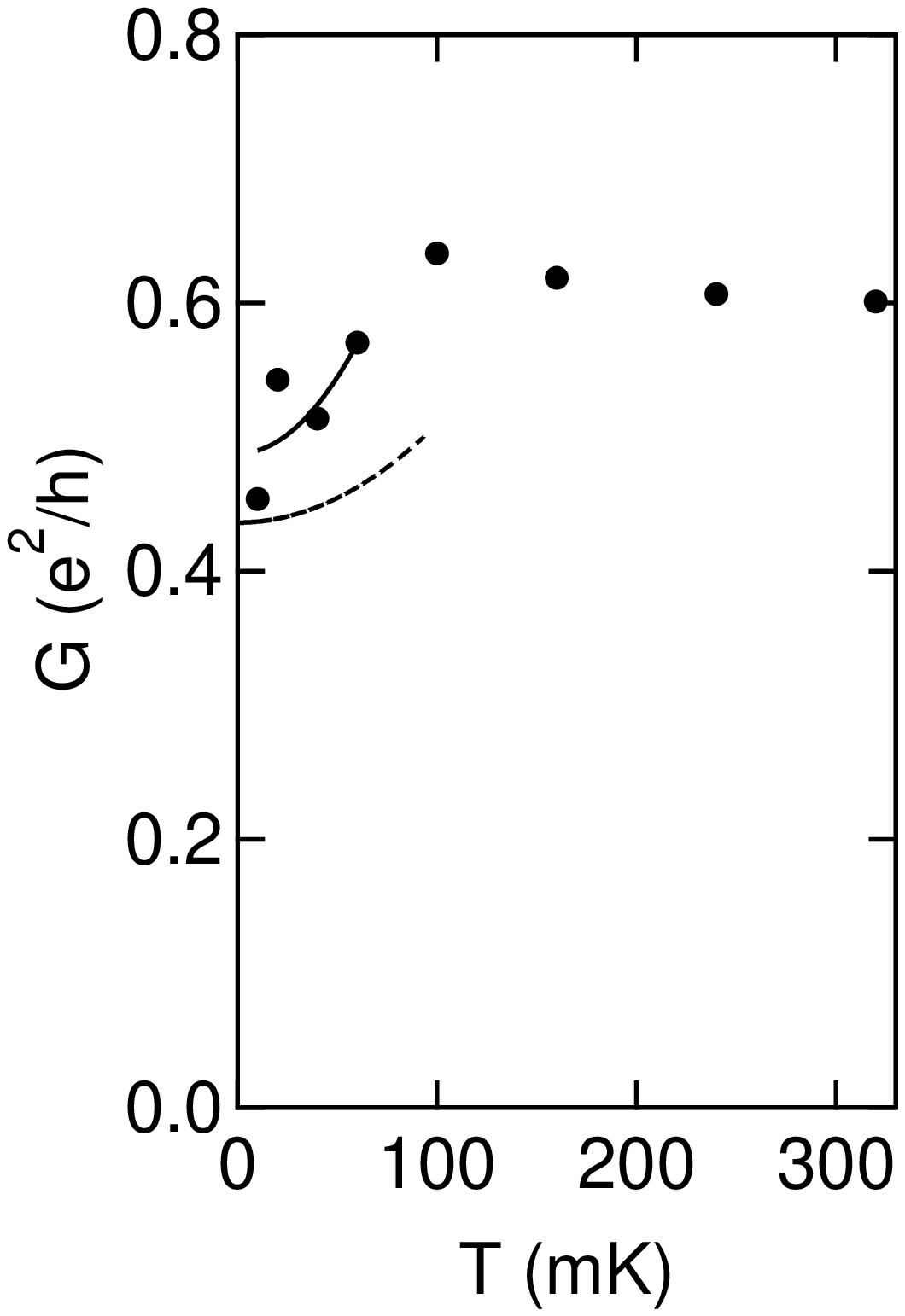}}
\put(0,5.2){(c)}
\put(0.5,0){\includegraphics[width=6.91cm, keepaspectratio=true]{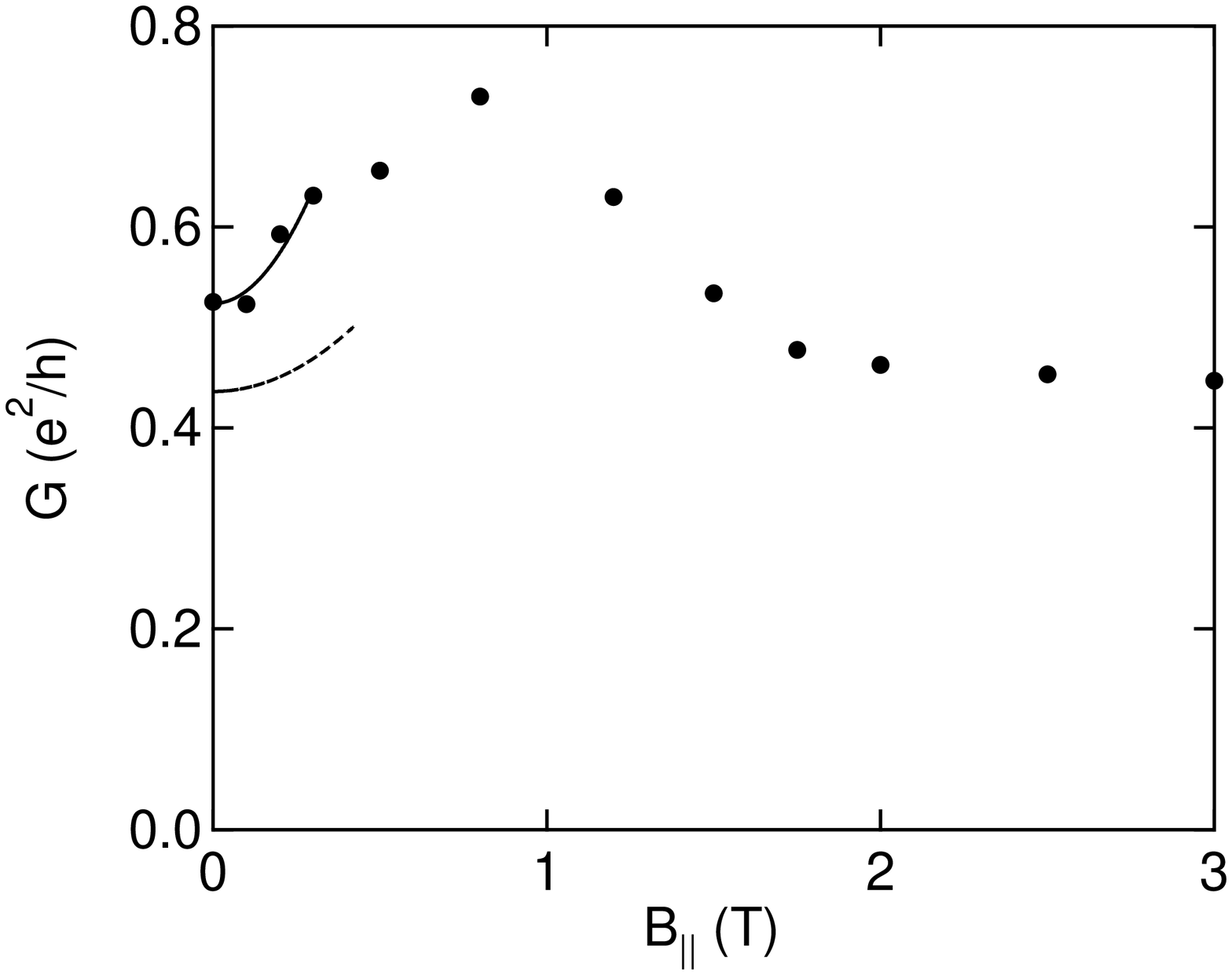}}
\end{picture}
\end{center}
\caption{(a) $\didv$ vs.~$\Vds$ at a constant \Vg=-595 mV (dots). A fit to Eq.~1 is shown near \Vds=0 (solid curve). (b) Mixing chamber temperature dependence of the zero-bias conductance G at \Vg=-595 mV (circles). The fit to Eq.~1 is also shown (solid curve). The dashed curve is the fit from (a) with the $\Vds$ axis converted into temperature. (c) $\Bpar$ dependence of G at \Vg=-595 mV (circles). The solid curve shows the fit to Eq.~1. The dashed curve is the fit from (a) with the $\Vds$ axis converted into magnetic field using $|g|=0.33$ (see text).}
\label{fig:FitsLowEnergy}
\end{figure}

Figure~\ref{fig:FitsLowEnergy} compares the dependences of $\didv$ on $\Vds$, \mbox{T}, and $\Bpar$ for a single value of $\Vg$. Figure~\ref{fig:FitsLowEnergy}(a) shows an expanded version of one of the \Bpar=0~T, base temperature curves in Fig.~\ref{fig:Diamond4}(b). As $\Vds$ is increased from zero, $\didv$ first increases and then decreases. A similar non-monotonic dependence on temperature is seen in the zero-bias conductance G, as shown in \subfig{FitsLowEnergy}{b}. Finally, the parallel magnetic field dependence of G is plotted in \subfig{FitsLowEnergy}{c}, and it also has a non-monotonic behavior. The non-monotonic dependence found for \didv(\Vds), G(\mbox{T}), and G(\Bpar) is in qualitative agreement with the prediction of Ref.~\cite{hofstetter2004:theoryST} for the two-stage Kondo effect that occurs when the triplet is screened by two modes. 

\section{Analysis}
\label{sec:analysis}

In order to extract the low energy scales from the dependences of $\didv$ on \Vds, \mbox{T}, and $\Bpar$ at fixed $\Vg$, we follow Ref.~\cite{hofstetter2004:theoryST}. Each dependence is predicted to be quadratic in lowest order. Summarizing all three dependences into one equation, the differential conductance in units of $e^2/h$ is predicted to depend on ($\Vds$,T,$\Bpar$) in the following way:

\begin{eqnarray}
&&\frac{\text{dI}}{\text{d}\Vds}(\Vds,\mbox{T},\Bpar) 
= G_0+ 
\label{eq:ParabolaST} 
\\
&&2(1-G_0)\left[
\left(\frac{\Vds}{\Vdsstar}\right)^2 + 
\left(\frac{\mbox{T}}{\Tstar}\right)^2 +
\left(\frac{\Bpar}{\Bstar}\right)^2
\right] \nonumber
\end{eqnarray} 

\noindent This equation has four parameters:  $G_0$ and ($\Vdsstar$,$\Tstar$,$\Bstar$). $G_0$ is the zero-bias conductance at the fixed value of $\Vg$ chosen. Any parameter among ($\Vdsstar$,$\Tstar$,$\Bstar$) is inversely proportional to the square root of the curvature of $\didv$ along the axis of the corresponding independent variable. The ``star'' parameters are simply related to the low energy scale of the two-stage Kondo system probed in three different ways.

In the experiments, only one quantity out of ($\Vds$,T,$\Bpar$) is varied at once. Therefore, for a given data set such as those in Fig.~\ref{fig:FitsLowEnergy}, the other two quantities are set to zero in Eq.~\ref{eq:ParabolaST}. This means this equation can be used to perform a two-parameter fit on the data points of $\didv(\Vds)$, G(T), or G(\Bpar). The mixing chamber temperature is nonzero and the electron temperature is even larger, so corrections to $G_0$ and T are expected. However, analysis shows that these corrections are in fact very small compared to the uncertainties in the data and the fitting procedure.

\begin{figure}[hbt] 
\setlength{\unitlength}{1cm}
\begin{center}
\begin{picture}(7,6)(0,0)
\put(0,0){\includegraphics[width=7.0cm, keepaspectratio=true]{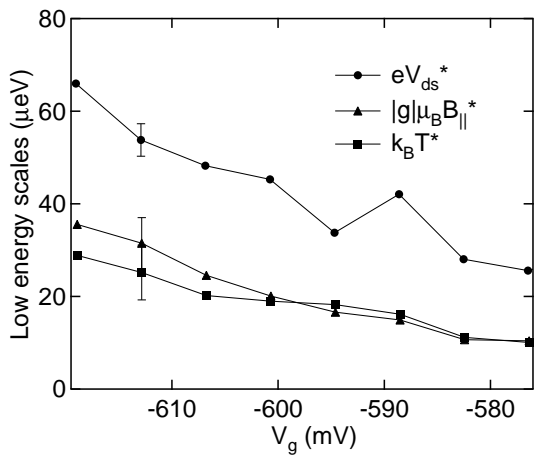}}
\end{picture}
\end{center}
\caption{Low energies $e\Vdsstar$ (circles), $|\mbox{g}|\mub\Bstar$ (triangles), and  $\kb\Tstar$ (squares) extracted from fits to Eq.~1 as a function of $\Vg$ for N=4. Representative error bars are shown.}
\label{fig:Energies}
\end{figure}

The fit of the form of Eq.~\ref{eq:ParabolaST} to data of G(\Bpar) at \Vg=-595 mV is shown as a solid curve in \subfig{FitsLowEnergy}{c}. The same type of fit can be repeated for all values of $\Vg$ in the Coulomb-blockade valley with N=4. These results are contained in Fig.~\ref{fig:Energies}. To our knowledge, it is the first time that measurements of $\Bstar$ are reported. Most authors who have studied similar systems have used a perpendicular magnetic field to access the triplet state.\cite{sasaki2000:verticalST,vdwiel2002:lateralST,zumbuhl2004:cotunnelingspec} A gate voltage serves this purpose here, which allows the use of a parallel magnetic field as a perturbation to measure $\Bstar$.

Eq.~\ref{eq:ParabolaST} can also be used to extract the low energy scales for the T and $\Vds$ dependences, $\Tstar$ and \Vdsstar, respectively.\cite{hofstetter2004:theoryST} For \Vg=-595 mV, the fit for \didv(\Vds) is plotted as a solid curve in \subfig{FitsLowEnergy}{a}, while the fit for G(T) is shown as a solid curve in \subfig{FitsLowEnergy}{b}. A standard conversion into energy is performed to compare $e\Vdsstar$, $\kb\Tstar$, and $|g|\mub\Bstar$ (with the g-factor $|g|=0.33$ and the Bohr magneton $\mub=58$~$\mu$eV/T). The results for the three scales are displayed in Fig.~\ref{fig:Energies} as a function of $\Vg$. The quantity $e\Vdsstar$ is significantly larger than $\kb\Tstar$ and $|g|\mub\Bstar$. The $\Vds\geq0$ portion of the fit in Fig.~\ref{fig:FitsLowEnergy}(a) is shown as a dashed curve in Figs.~\ref{fig:FitsLowEnergy}(b) and (c) as another illustration of this difference.

The value $|g|=0.33\pm0.02$ used to convert $\Bstar$ into energy is extracted at $\Vg=-637$~mV, i.e.~for N=3 and $S=1/2$,  from the slope of the splitting of the spin-flip inelastic cotunneling threshold versus $\Bpar$ up to 3~T (not shown). This method is described in more detail elsewhere.\cite{kogan2004:gfactor} This value of $|g|$ agrees well with the value $|g|=0.32\pm0.02$ extracted from another sample made on the same heterostructure with an odd number of electrons for $\Bpar$ up to 5~T. The value $|g|=0.33\pm0.02$ from the N=3 case is used for the conversion of $\Bstar$ into energy.

\section{Discussion and Conclusions}
\label{sec:discussion}

From Fig.~\ref{fig:Energies} one sees that the low energy scales for the temperature and $\Bpar$ dependences agree very well, as expected.\cite{hofstetter2004:theoryST} However, the scale determined from the voltage dependence is significantly larger. The scale $e\Vdsstar$ is obtained by applying a finite source-drain voltage, which takes the system out of equilibrium. In contrast, $|g|\mub\Bstar$ and $\kb\Tstar$ are extracted in equilibrium from the zero-bias conductance. Nonequilibrium effects may therefore be the cause of this discrepancy.

The low energy scales $E_l$ extracted from our measurements vary roughly by a factor three over the range of $\Vg$ displayed in Fig.~\ref{fig:Energies}. Hofstetter and Zarand~\cite{hofstetter2004:theoryST} predict that $E_l$ decreases exponentially as the energy spacing $\Delta$ between the orbitals decreases when $\Delta$ is less than $2E_S$, where $E_S$ is the (ferromagnetic) exchange energy. Thus one expects the exponential dependence when the triplet is the ground state.

Figure~\ref{fig:Singlet-triplet}(f) shows the transition from singlet to triplet, very similar to those studied by Kogan \etal\cite{kogan2003:ST_zeroB} These authors have extracted $E_S=(0.08\pm0.04)$~meV from a situation where two transitions, into and out of the triplet, occur within one Coulomb-blockade diamond, thanks to the change in dot shape induced by a gate electrode. In the present experiment, the sensitivity of the singlet-triplet excitation energy to $\Vg$ (on the singlet side of the transition) is smaller than in that of Kogan \etal The slope of the inelastic cotunneling feature corresponding to the triplet excited state in the $\Vds$-$\Vg$ plane in Fig.~\ref{fig:Singlet-triplet}(a) is $6.39\times10^{-3}\,e$, which is more than three times smaller than the value $1.95\times10^{-2}\,e$ found by Kogan \etal However, Kogan \etal have reported slopes in the range $5\times10^{-3}\,e$ to $2\times10^{-2}\,e$, depending on the voltage configuration on the confinement electrodes. This implies that the sensitivity of the singlet-triplet excitation energy depends not only on the geometry of the electrodes, but also on the exact dot shape created by the energized electrodes. 

In our experiment, we observe only one transition, from the singlet to the triplet, and not the one back to the singlet; therefore, we cannot determine $E_S$. Examining the singlet-triplet excitation energy as a function of $\Vr$ reveals a dependence that is approximately linear in $\Vr$ away from the transition. This allows us to put a lower bound on the exchange energy of $E_S\geq0.15$~meV. Nonetheless, the data of Fig.~\ref{fig:Singlet-triplet}(f) make it clear that, for the experiments in Fig.~\ref{fig:Energies}, the triplet is the ground state and that the exponential dependence of $E_l$ on $\Delta$ is expected.

We note that two energy scales in the Kondo effect have also been predicted for double dots. (See Cornaglia and Grempel\cite{cornaglia2005:DQDD} and references therein.) However, in double dots the exchange is antiferromagnetic, whereas in single dots, like ours, it is ferromagnetic, so the mechanism leading to the two-stage Kondo effect is different.

The two-level model used by Kogan \etalComma~which includes a mixing between the orbitals proportional to the change in $\Vg$, predicts that the two orbitals repel each other and undergo an avoided crossing. 
Such an avoided crossing is also expected in our experiments, and it causes $\Delta$ to vary quadratically with gate voltage near its minimum. The slow variation of $E_l$ with $\Vg$ may occur because the avoided crossing prevents $\Delta$ from decreasing to zero. Thus, even if $E_l$ were exponential in $\Delta$, it would not be exponential in $\Vg$. It is also possible that the symmetry between the two modes leads to saturation,\cite{hofstetter2004:theoryST} and this effect is yet another candidate to explain the slow variation of $E_l$ with $\Vg$.

In summary, a lateral quantum dot with four electrons has been driven through a singlet-triplet crossover. Behavior consistent with a two-stage Kondo description is found for the triplet. The lower energy scales $\Bstar$, $\Vdsstar$, and $\Tstar$ have been extracted. It is found that $\kb\Tstar$ is in good quantitative agreement with $|g|\mub\Bstar$, but that $e\Vdsstar$ is larger, probably because of nonequilibrium effects. 

\acknowledgments

We are grateful to W.~Hofstetter, S.~Amasha, and L. Levitov for discussions and to D.~M.~Zumb\"{u}hl, L.~E.~Calvet, A.~Wang, O.~Dial, and G.~Steele for experimental help. We thank R.~C.~Ashoori for sharing his photolithography facilities. G.G. acknowledges partial support from the Natural Sciences and Engineering Research Council of Canada. This work was supported by the US Army Research Office under Contract DAAD19-01-1-0637, by the National Science Foundation under Grant No.~DMR-0353209, and in part by the NSEC Program of the National Science Foundation under Award Number DMR-0117795.

\bibliography{granger_2SK}

\end{document}